\documentstyle[12pt]{article}

\skewchar\fivmi='177
\skewchar\sixmi='177
\skewchar\sevmi='177
\skewchar\egtmi='177
\skewchar\ninmi='177
\skewchar\tenmi='177
\skewchar\elvmi='177
\skewchar\twlmi='177
\skewchar\frtnmi='177
\skewchar\svtnmi='177
\skewchar\twtymi='177
\def\@magscale#1{ scaled \magstep #1}
\skewchar\fivsy='60
\skewchar\sixsy='60
\skewchar\sevsy='60
\skewchar\egtsy='60
\skewchar\ninsy='60
\skewchar\tensy='60
\skewchar\elvsy='60
\skewchar\twlsy='60
\skewchar\frtnsy='60
\skewchar\svtnsy='60
\skewchar\twtysy='60

\catcode`@=11
\def\un#1{\relax\ifmmode\@@underline#1\else
        $\@@underline{\hbox{#1}}$\relax\fi}
\catcode`@=12

\def\a{\alpha}
\def\b{\beta}

\def\d{\delta}

\def\g{\gamma}

\def\l{\lambda}

\def\s{\sigma}

\def\L{\Lambda}

\def\dslash{\not{\hbox{\kern-2pt $\partial$}}}
\def\Dslash{\not{\hbox{\kern-4pt $D$}}}
\def\pslash{\not{\hbox{\kern-2.3pt $p$}}}
 \newtoks\slashfraction
 \slashfraction={.13}
 \def\slash#1{\setbox0\hbox{$ #1 $}
 \setbox0\hbox to \the\slashfraction\wd0{\hss \box0}/\box0 }

\font\ro=cmsy10                          
\def\kcr{{\hbox{\ro \char'170}}}                
\def\ktl{{\hbox{\ro \char'170}}}        
\def\ktr{{\hbox{\ro \char'170}}}        
\def\kbl{{\hbox{\ro \char'170}}}        
\def\kbr{{\hbox{\ro \char'170}}}        

\def\plpl{\raise-2pt\hbox{$\raise3pt\hbox{$_+$}\hskip-6.67pt\raise0.0pt
\hbox{$^+$}\hskip 0.01pt$}}
\def\mimi{\raise-2pt\hbox{$\raise3pt\hbox{$_-$}\hskip-6.67pt\raise0.0pt
\hbox{$^-$}\hskip 0.01pt$}}

\def\bo{{\raise.15ex\hbox{\large$\Box$}}}               
\def\pa{\partial}                                       
\def\TH{{\raise.2ex\hbox{$\displaystyle \bigodot$}\mskip-4.7mu \llap H \;}}
\def\face{{\raise.2ex\hbox{$\displaystyle \bigodot$}\mskip-2.2mu \llap {$\ddot
        \smile$}}}                                      

\def\sp#1{{}^{#1}}                              
\def\sb#1{{}_{#1}}                              
\def\leftrightarrowfill{$\mathsurround=0pt \mathord\leftarrow \mkern-6mu
        \cleaders\hbox{$\mkern-2mu \mathord- \mkern-2mu$}\hfill
        \mkern-6mu \mathord\rightarrow$}
\def\dvec#1{\vbox{\ialign{##\crcr
        \leftrightarrowfill\crcr\noalign{\kern-1pt\nointerlineskip}
        $\hfil\displaystyle{#1}\hfil$\crcr}}}           

\def\fracm#1#2{\hbox{\large{${\frac{{#1}}{{#2}}}$}}}
\def\frac#1#2{{\textstyle{#1\over\vphantom2\smash{\raise.20ex
        \hbox{$\scriptstyle{#2}$}}}}}                   
\def\sfrac#1#2{{\vphantom1\smash{\lower.5ex\hbox{\small$#1$}}\over
        \vphantom1\smash{\raise.4ex\hbox{\small$#2$}}}} 
\def\bfrac#1#2{{\vphantom1\smash{\lower.5ex\hbox{$#1$}}\over
        \vphantom1\smash{\raise.3ex\hbox{$#2$}}}}       
\def\afrac#1#2{{\vphantom1\smash{\lower.5ex\hbox{$#1$}}\over#2}}    

\newskip\humongous \humongous=0pt plus 1000pt minus 1000pt

\newif\ifdtup

\def\ref#1{$\sp{#1)}$}

\parskip=\medskipamount                 
\thicklines                         

\def\oldheadpic{                                
        \setlength{\unitlength}{.4mm}
        \thinlines
        \par
        \begin{picture}(349,16)
        \put(325,16){\line(1,0){4}}
        \put(330,16){\line(1,0){4}}
        \put(340,16){\line(1,0){4}}
        \put(335,0){\line(1,0){4}}
        \put(340,0){\line(1,0){4}}
        \put(345,0){\line(1,0){4}}
        \put(329,0){\line(0,1){16}}
        \put(330,0){\line(0,1){16}}
        \put(339,0){\line(0,1){16}}
        \put(340,0){\line(0,1){16}}
        \put(344,0){\line(0,1){16}}
        \put(345,0){\line(0,1){16}}
        \put(329,16){\oval(8,32)[bl]}
        \put(330,16){\oval(8,32)[br]}
        \put(339,0){\oval(8,32)[tl]}
        \put(345,0){\oval(8,32)[tr]}
        \end{picture}
        \par
        \thicklines
        \vskip.2in}
\def\oldtitle#1#2#3#4{\oldheadpic\begin{center}\vglue.5in{\large\bf #1}\\[.6in]
        {#2}\\[.1in] {\it Department of Physics and Astronomy}\\
        {\it University of Maryland, College Park, MD 20742}\\[.6in]
        Physics Publication \#{#3}\\ {#4}\\[1.5in] {\bf ABSTRACT}\\[.1in]
        \end{center} \begin{quotation}}                 
\def\oldTitle#1#2#3#4#5#6#7{\oldheadpic\begin{center} \vglue .4in
        {\large\bf #1}\\[.4in]
        {#2}\\[.1in] {\it Department of Physics and Astronomy}\\
        {\it University of Maryland, College Park, MD 20742}\\[.1in]
        {#3}\\[.1in] {\it {#4}}\\ {\it {#5}}\\[.4in]
        Physics Publication \#{#6}\\ {#7}\\[.5in] {\bf ABSTRACT}\\[.1in]
        \end{center} \begin{quotation}}                 
\def\border{                                            
        \setlength{\unitlength}{1mm}
        \newcount\xco
        \newcount\yco
        \xco=-21
        \yco=12
        \begin{picture}(140,0)
        \put(\xco,\yco){$\ktl$}
        \advance\yco by-1
        {\loop
        \put(\xco,\yco){$\kcr$}
        \advance\yco by-2
        \ifnum\yco>-240
        \repeat
        \put(\xco,\yco){$\kbl$}}
        \xco=158
        \yco=12
        \put(\xco,\yco){$\ktr$}
        \advance\yco by-1
        {\loop
        \put(\xco,\yco){$\kcr$}
        \advance\yco by-2
        \ifnum\yco>-240
        \repeat
        \put(\xco,\yco){$\kbr$}}
        \put(-20,13){\tiny University of Maryland Elementary Particle
Physics University of Maryland Elementary Particle Physics University of
Maryland Elementary Particle Physics}
        \put(-20,-241.5){\tiny University of Maryland Elementary
Particle Physics University of Maryland Elementary Particle Physics
University of Maryland Elementary Particle Physics}
        \end{picture}
        \par\vskip-8mm}
\def\bordero{                                           
        \setlength{\unitlength}{1mm}
        \newcount\xco
        \newcount\yco
        \xco=-31
        \yco=12
        \begin{picture}(140,0)
        \put(\xco,\yco){$\ktl$}
        \advance\yco by-1
        {\loop
        \put(\xco,\yco){$\kclr}
        \advance\yco by-2
        \ifnum\yco>-240
        \repeat
        \put(\xco,\yco){$\kbl$}}
        \xco=151
        \yco=12
        \put(\xco,\yco){$\ktr$}
        \advance\yco by-1
        {\loop
        \put(\xco,\yco){$\kcr$}
        \advance\yco by-2
        \ifnum\yco>-240
        \repeat
        \put(\xco,\yco){$\kbr$}}
        \put(-20,12){\ooo
bacdefghidfghghdhededbihdgdfdfhhdheidhdhebaaahjhhdahba

hgdedge
   hgfdiehhgdigicba}
        \put(-20,-241.5){\ooo
ababaighefdbfghgeahgdfgafagihdidihiidhiagfedhadbfd

ecdcdfa
   gdcbhaddhbgfchbgfdacfediacbabab}
        \end{picture}
        \par\vskip-8mm}
\def\headpic{                                           
        \indent
        \setlength{\unitlength}{.4mm}
        \thinlines
        \par
        \begin{picture}(29,16)
        \put(165,16){\line(1,0){4}}
        \put(170,16){\line(1,0){4}}
        \put(180,16){\line(1,0){4}}
        \put(175,0){\line(1,0){4}}
        \put(180,0){\line(1,0){4}}
        \put(185,0){\line(1,0){4}}
        \put(169,0){\line(0,1){16}}
        \put(170,0){\line(0,1){16}}
        \put(179,0){\line(0,1){16}}
        \put(180,0){\line(0,1){16}}
        \put(184,0){\line(0,1){16}}
        \put(185,0){\line(0,1){16}}
        \put(169,16){\oval(8,32)[bl]}
        \put(170,16){\oval(8,32)[br]}
        \put(179,0){\oval(8,32)[tl]}
        \put(185,0){\oval(8,32)[tr]}
        \end{picture}
        \par\vskip-6.5mm
        \thicklines}
\def\title#1#2#3#4{\border\headpic {\hbox to\hsize{#4 \hfill UMDEPP #3}}\par
        \begin{center} \vglue .5in {\large\bf #1}\\[.6in]
        {#2}\\[.1in] {\it Department of Physics and Astronomy}\\
        {\it University of Maryland, College Park, MD 20742}\\[1.5in]
        {\bf ABSTRACT}\\[.1in] \end{center} \begin{quotation}}  
\def\Title#1#2#3#4#5#6#7{\border\headpic
        {\hbox to\hsize{#7 \hfill UMDEPP #6}}\par
        \begin{center} \vglue .4in {\large\bf #1}\\[.4in]
        {#2}\\[.1in] {\it Department of Physics and Astronomy}\\
        {\it University of Maryland, College Park, MD 20742}\\[.1in]
        {#3}\\[.1in] {\it {#4}}\\ {\it {#5}}\\[.5in] {\bf ABSTRACT}\\[.1in]
        \end{center} \begin{quotation}}                 
\def\endtitle{\end{quotation}\newpage}                  

\begin{document}

\def\fracm#1#2{\hbox{\large{${\frac{{#1}}{{#2}}}$}}}
\def\gfrac#1#2{\frac {\scriptstyle{#1}}
        {\mbox{\raisebox{-.6ex}{$\scriptstyle{#2}$}}}}
\def\gg{{\hbox{\sc g}}}
\border\headpic {\hbox to\hsize{
July 1994 \hfill {UMDEPP 95-06}}}

\par
\setlength{\oddsidemargin}{0.3in}
\setlength{\evensidemargin}{-0.3in}
\begin{center}
\vglue .08in
{\large\bf Why Are There So Many N = 4 Superstrings?}
\\[.72in]

S. James Gates, Jr.
\footnote {Supported by National Science Foundation Grant PHY-91-19746}${}^,
$\footnote {Supported by NATO Grant CRG-93-0789}\\[.02in]
{\it Department of Physics\\
University of Maryland\\
College Park, MD 20742-4111  USA}\\
{\bf {\tt gates@umdhep.umd.edu}}\\[1.2in]

{\bf ABSTRACT}\\[.002in]
\end{center}
\begin{quotation}
{We demonstrate the existence of three off-shell distinct N = 4 superstrings
at the level of manifest locally supersymmetric Lagrangian field theories
on the world sheet.}

\endtitle
\section{Introduction}
\indent

There are three 2D, N = 4 off-shell scalar multiplets with a finite
number of auxiliary fields.  This is a little known fact that is
the primary motivation for this presentation.  In the following we
will give, to our knowledge, the first comprehensive discussion regarding
this curious situation. The fact that this situation exist at all
is extremely puzzling.  There is a known precedent for this
situation. The resolution of the puzzle in the precedent N = 2 models
\cite{GaLO} resides in the existence of ``mirror symmetry.''  The four
distinct off-shell N = 2 models pointed out in reference \cite{GaLO} are
related pair-wise by a mirror transformation that replaces chiral superfields
by twisted chiral superfields.  The existence of mirror symmetry,
at least at the level of lagrangian field theories, is a consequence that
follows since both chiral and twisted chiral multiplets \cite{GaHR} exist as
distinct off-shell N = 2 scalar multiplet representations.  As we argued,
N = 2 superstrings can be constructed utilizing two chiral multiplets,
two anti-chiral multiplets and one chiral multiplet and one twisted
chiral multiplet. Interestingly enough, models of this last type have
recently appeared in the context of exact solutions for ``stringy'' plane
waves \cite{KKL}.

For a long time, issues regarding N = 4 superstrings have simply not
seemed pressing. Apart from the original work discovering these theories
\cite{11IJHS}, there has been a little work carried out on their
quantization \cite{vanN}.  This was largely due to the belief that the
critical dimension of such theories was minus one.  This ``standard folklore''
has recently been challanged \cite{sieg}, with the proposal that the
actual critical value is plus one!  Even should this new proposal not
stand the test of time, the recent new insight into integrable supersymmetric
models that can be understood as string theories away from their critical
dimensions suggest that there is reason to believe that interesting
systems may very well be described by N = 4 superstring theories. Fortunately,
we are very well situated to study N = 4 superstrings. Off-shell structures
have been known for some time for 2D, N = 4 supergravity \cite{GaLO,GHvN}.

\section{Type-I and Type-II, N = 4 Superstrings}
\indent

These are theories of N = 4 strings that are based on twisted N = 4 scalar
multiplets.  These types of scalar multiplets may be regarded as having
origins as dimensional reductions from 3D or 4D. There are 8 - 8 fermion
and boson degrees of freedom in these scalar multiplets.

We start with a 4D, N = 2 vector multiplet and a 4D, N = 2 tensor
multiplet.  Considering a compactification to 3D splits off some of
the components of the gauge fields in each multiplet into different
3D representations of the SO(1,2) group.  The 4D vector gauge field
``yields'' a 3D vector gauge field and one scalar.  Similarly, a 4D,
2-form gauge field ``yields'' a 3D, 2-form gauge field
and a 3D vector gauge field.  In 3D, by a duality
transformation, the former can be replaced by an auxiliary scalar.
Since an irreducible spinor representation of SO(1,3) contains
two irreducible spinor representations of SO(1,2) The number of
supersymmetries for the multiplets double. So the multiplets that
result are 3D, N = 4 representations.  We conclude there may exist
two distinct 3D, N = 4 vector multiplets\footnote{The existence
of these two distinct 3D, N = 4 vector multiplets plays a very
important role in topological field theory \cite{BrkGT}.}.  These
two different representations of the same physical degrees of freedom
are an example of the occurence of variant superfield representations
\cite{SG1}.

Now we continue from 3D to 2D via a further compactification.
This time the 3D vector gauge field in each multiplet ``yields''
a 2D vector gauge field as well as one scalar. In 2D, a gauge vector
propagates no physical degrees of freedom and by a duality transformation
can be replaced by an auxiliary scalar.  The 4D, N = 2 vector
multiplet reduces to a 2D, N = 4 scalar multiplet (called the twisted-I
multiplet) and the 4D, N = 2 tensor multiplet also reduces to a 2D, N = 4
scalar multiplet (called the twisted-II multiplet). The explicit form of
these two 2D, N = 4 twisted scalar multiplets are given below.

The twisted-I multiplet has been discussed several times before
\cite{GaLO,GaHR}. The supersymmetry
variations are characterized by,
$$\begin{array}{lll}
D_{\a i} F  &=& 2  C_{i j} \psi_{\a} {}^j ~~~, \\

{\bar D}_{\a}{}^{ i} F  &=& 0 ~~~, \\

D_{\a i} S  &=& -i  {\bar \psi}_{\a i}  ~~~, \\

D_{\a i} P  &=&  (\g^3)_{\a} {}^{\b} {\bar \psi }_{\b i} ~~~, \\

D_{\a i} \psi^{\b j} &=&  \d_i {}^j \left [~ (\g^c)_\a
{}^{\b} (\pa_c S) ~+~ i(\g^3 \g^c)_\a  {}^{\b} (\pa_c P)~\right ]  \\
& &+ ~ \fracm 12  \left [ ~\d_i {}^j (\g^3)_{\a} {}^{\b} A ~+~ i \d_{\a }
{}^{\b} A_i {}^j ~\right ] ~~, \\

{\bar D}_{\a}{}^{i} \psi^{\b j} &=& i C^{i j}
(\g^c)_\a  {}^{\b} (\pa_c F) ~~, \\

D_{\a i} A &=& -i 2  (\g^3 \g^c)_{\a} {}^{\b} {\pa}_c
{\bar \psi }_{\b i}  ~~, \\
\end{array}    $$
\begin{equation}
D_{\a i} A_j {}^k  = {~~}4 (\d_j {}^l \d_i {}^k - \frac 12
\d_j {}^k \d_i {}^l ) (\g^c)_{\a} {}^{\b} {\pa}_c {\bar \psi
}_{\b l}  ~~, {~~~~~}
\end{equation}
With the exception of $F$ and ${\psi}_{\a i}$, all the remaining fields
are real (for $ A_i {}^j = ( A_j {}^i)^* $).

The invariant component level action takes the explicit form,
$$
{\cal S}_{{\rm T}{\rm M}{\rm {- I}} ,{\rm N} = 4} ~=~  \int d^2
\s ~ [~  \frac 12 S \bo S ~+~  \frac 12 P \bo P ~+~
 \frac 12 F \bo  {\bar F} ~+~ i
{ \psi}^{\a i} ( \g \sp{c})  \sb{\a \b} \pa_c
{\bar \psi}^{\b}{}_{i} {~~~~~~} {~~~~~~}
$$
\begin{equation}
{~~~~~~~~~~}  -  \frac 14 A^2 ~-~  \frac 1{16} A_i {}^j A_j {}^i ~ ~ ] ~~.
{~~~~~~}
\end{equation}

In the presence of 2D, N = 4 supergravity, we can write an action for
this multiplet if we introduce two pre-potential superfields; $V$ and
$V_i {}^j$. The highest components of $V$ and $V_i {}^j$ correspond to
$A$ and $A_i {}^j$, respectively.  The resulting expression for the action
describes the  twisted-I, N = 4 superstring prior to any gauge fixing.
\begin{equation}
{~~~~~~~~~~}  {~~~~~~~~~~} {\cal S}_{{\rm T}{\rm M}{\rm {- I}} ,{\rm N}
= 4} ~=~ - \int d^2 \s d^8 \theta~ E^{-1} ~[~ V  A ~+~ V_i {}^j A_j {}^i ~]
 ~~~. {~~~~~~~~~~}  {~~~~~~~~~~}
\end{equation}

The twisted-II theory was first found by Ivanov and
Krivonos \cite{IVK}.  A description that is equivalent to their
previous work is given by,
$$~~~~~~~~~~~~~{D}_{\a i } {\cal T}  ~=~
(\g \sp{3})  \sb{\a} \sp{\b} {\Psi}_{\b i}  ~~~,
{}~~~~~~~~~~~~~~~~~~~~~~~~~~~~~~~~~~~~~~~~~~~~~~~~~~$$
$$~~~~~~~~~~~~~~~~~~~~~{D}_{\a i } {\cal X} {}_j {}^k ~=~   i \left [
\d_i {}^k { \Psi}_{\a j} ~-~ \fracm 12 \d_j {}^k {\Psi}_{\a i}  \right ] ~~~,
{}~~~~~~~~~~~~~~~~~~~~~~~~~~~~~~~~~~~~~~~~~~~~~~~~$$
$$~~~~~~~~~~~~~~~~~~~~~~~~~~~~~~ {\cal X} {}_i {}^i  ~=~ 0 ~~~, ~~~ {\cal X}
{}_i {}^j ~-~ ( {\cal X} {}_j {}^i )^* ~=~ 0 ~~~,
{}~~~~~~~~~~~~~~~~~~~~~~~~~~~~~~~~~~~~~~~~~~~~~~~~~~~$$
$$~~~~~~~~~~~~~{D}_{\a i } {\Psi}_{\b j}  ~=~ \fracm 12  C_{\a \b} C_{ i j}
{\bar J}~~~, ~~~~~~~~~~~~~~~~~~~~~~~~~~~~~~~~~~~~~~~~~~~~~~~~~~~~~~$$
$$~~~~~~~~~~~~~~~~~~~~~~~~~~~~{D}_{\a i } {\bar J} ~=~ 0 ~~~,~~~ m ~-~
({m} )^* ~=~ 0 ~~~, n ~-~ ({n} )^* ~=~ 0 ~~~, ~~~~~~~~~~~~~~~~~~~~~~~~~~~~$$
$${\bar D}_{\a}{}^{ i } {\Psi}_{\b j}  ~=~ i \d_j {}^i
(\g \sp{3} \g \sp{a})  \sb{\a \b}  \left ( \pa_a {\cal T} \right ) ~+~ 2
(\g \sp{a})  \sb{\a \b} \left ( \pa \sb{a} {\cal X}_{j}{}^i \right )
{}~~~~~~~$$
$$~~~~~~~~~~~~~
 ~+~ i \fracm 12 C_{\a \b} \d \sb j {}^i m ~+~ \fracm 12 (\g \sp{3})
\sb{\a \b}  \d \sb j {}^i n ~~. ~~~~~~~~~~$$
$$ ~~~~~{D}_{\a i } { J} ~=~ - i 4  C_{ i j} (\g \sp{a})  \sb{\a} \sp{\b}
\left ( \pa_a  {\bar \Psi}_{\b}{}^{j} \right ) ~~~,~~~~~~~~~~~~~~
{}~~~~~~~~~~~ $$
$$ ~~~{D}_{\a i } n ~=~ - i 2 (\g \sp{3} \g \sp{a})  \sb{\a} \sp{\b}
\left ( \pa_a  { \Psi}_{\b i} \right ) ~~~,~~~~~~~~~~~~~~~~~~~
{}~~~~~~~~ $$
\begin{equation}
{}~{D}_{\a i } m ~=~ - 2 ( \g \sp{a})  \sb{\a} \sp{\b}
\left ( \pa_a  { \Psi}_{\b i} \right ) ~~~.~~~~~~~~~~~~~~~~
{}~~~~~~~~~~~~~
\end{equation}
Here the complex fields are $J$ and ${\Psi}_{\a i}$.

An invariant component level action takes the explicit
form,
$$
{\cal S}_{{\rm T}{\rm M}{\rm {- II}} ,{\rm N} = 4} ~=~  \int d^2
\s ~ [~  \frac 12  {\cal T}  \bo  {\cal T} ~+~
 {\cal X} {}_j {}^i  \bo  {\cal X} {}_i {}^j ~+~ i
{ \Psi}^{\a}{}_{i} ( \g \sp{c})  \sb{\a \b} \pa_c
{\bar \Psi}^{\b}{}^{i} {~~~~~~} {~~~~~~}
$$
\begin{equation}
{~~~~} {~~~~}  ~-~
\frac 18 (~ m^2 ~+~ n^2 ~+~ {J}{\bar J} ~) ~ ] ~~.  {~~~~~~}
\end{equation}

In the presence of 2D, N = 4 supergravity, we can write an action for
this multiplet if we introduce three pre-potential superfields;
$K$, $L$ and $\L$ (with ${\bar D}_{\a}{}^i \L = 0$). The highest
components of $K$, $L$ and $\L$ correspond to $m$, $n$ and $J$,
respectively.  The resulting expression for the action describes the
twisted-II, N = 4 superstring prior to any gauge fixing,
$$
{\cal S}_{{\rm T}{\rm M}{\rm {- II}} ,{\rm N} = 4} ~=~ - \int d^2
\s d^8 \theta~ E^{-1} ~[~ K m ~+~ L n ~]  {~~~~~~~~~~}  {~~~~~~~~~~}
{~~~~~~~~~~} {~~~~~~~~~~} $$
\begin{equation}
{~~~~} {~~~~}  ~-~ \int d^2 \s d^4 \theta~ {\cal E}^{-1} ~ \L {J} ~-~
 \int d^2 \s d^4 {\bar \theta}~ {\bar {\cal E}}^{-1} ~ {\bar \L}
{\bar J} ~~  ~~.  {~~~~~~}
\end{equation}
Here ${\cal E}^{-1}$ and ${\bar {\cal E}}^{-1}$ denote chiral and
anti-chiral density measures.

Even in 2D these two multiplets possess a ``duality.''   In terms of
equations, this duality is made explicit by observing,
$$
D_{\a i} F + i 2 C_{i j} {\bar D}_{\a}{}^j S = 0 ~~, ~~ {\bar D}_{\a}{}^i
F = 0 ~~,~~  D_{\a i} S + i (\g^3)_{\a}{}^{\b}  D_{\b i} P = 0 ~~~~, $$
\begin{equation}
D_{\a i} J + i 2 C_{i j} {\bar D}_{\a}{}^j m = 0 ~~, ~~ {\bar D}_{\a}{}^i
J = 0 ~~,~~  D_{\a i} m + i (\g^3)_{\a}{}^{\b}  D_{\b i} n = 0 ~~~~.
\end{equation}
That is, we can define a duality map that interchanges the physical
spin-0 fields in one of the twisted multiplets with the auxiliary
spin-0 fields in the other one of the twisted multiplets and vice-versa.

The most powerful manifestation of this phenomena is the existence of an
N = 4 mass term!  Previously the only known way to introduce such mass
terms in 2D, N = 4 theories required the use of ``active central charges''
that was discussed a very long time ago \cite{FrdAlGat}.  Since these two
dual N = 4 scalar multiplets exist, there is a much simpler way to construct
massive 2D, N = 4 representations.  This mass is produced in ``associated
production.''  That is, a Twisted-I multiplet and a Twisted-II multiplet
simultaneously acquire mass.  This phenomena of the associated production
of mass has been observed before \cite{DOGT} in the context of 4D, N = 1
theories. Here it is easy to show that the following action is invariant
under the supersymmetry variations described above.
$$
{\cal S}_{{\rm N} = 4, {\rm {mass}}} ~=~  \int d^2 \s ~ M_0 [ ~
  \frac 12 m S ~-~  \frac 12 n P ~-~ \frac 14 (~ J {\bar F} ~+~
{\bar J} F~)   {~~~~~~~~~~~~~~~~~~~~~}
$$
\begin{equation}
{~~~~~~~} {~~~~~~~} {~~~~~~~~~~~~~~~} -~ \frac 18 {\cal X} {}_i {}^j
A_j {}^i ~-~
\frac 12 {\cal T}
A ~+~ (~ {\bar \Psi}^{\a i} {\bar \psi}_{\a i} ~+~ {\rm h.} {\rm c.}
 ~) ~] ~~~.
\end{equation}
We should mention perhaps that this mass term is to be expected from
the point of view of a recently derived result for 3D, N = 4
BF theory \cite{BrkGT}.

\section{The HM-II, N = 4 Superstring}
\indent

Some time ago, an off-shell scalar multiplet (called the ``relaxed
hypermultiplet'') with manifest 4D, N = 2 supersymmetry was found
\cite{HoSTT}. This led to the discovery of a whole previously unknown
class of finite N = 2 supersymmetric Yang-Mills-matter models \cite{WePK}.
This off-shell representation also gives a very simple tool
to derive yet one more off-shell formulation of a superstring with 2D,
N = 4 supersymmetry.  Only a simple toroidal compactification of the
4D action is required!  The number of components in this scalar multiplet
is far greater than in the twisted type-I and type-II twisted scalar
multiplets. It has 32-32 fermionic and bosonic degrees of freedom.
We will call the resultant N = 4 superstring the HM-II, N = 4 superstring.
Since the number of fields is so large, it is simplest to begin with
its superspace action.
\begin{equation}
 {\cal S}_{{\rm HM},{\rm N} = 4} ~=~  \int d^2 \s d^8
\theta ~ E^{-1} {~} [ ~ (\lambda_{\a}{}^i \rho^{\a} {}_i ~+~
{\bar \lambda}_{\a}{}_i {\bar \rho}^{\a} {}^i ) ~+~ L^{ijkl} X_{ijkl}
{}~] ~~~.
\end{equation}

In this expression the fundamentally unconstrained superfield
potentials are $ \rho^{\a} {}_i $ and $X_{ijkl}$ (with $X_{ijkl}
= C_{i i'} C_{j j'} C_{k k'} C_{l l'} {X^{i'j'k'l'}}^*$).  In this action,
the superdeterminant density is that associated with the minimal N = 4
supergravity theory described in \cite{GaLO}.

In order to describe this theory at the level of components, we simply
``borrow'' the results from reference \cite{HoSTT}.  However, we first
make certain field re-definitions that ``diagonalize'' the component
action and result in the obvious separation of auxiliary fields
from propagating fields. These re-definitions take the forms,
$$
\varphi \equiv [V] ~~~,~~~ \varphi_{i j}  \equiv  2 [L_{i j}] ~~~, ~~~ \psi^{\a
i}
 \equiv  \frac 43 [\l^{\a i }] ~~~, $$
$$ \l^{\a i} \equiv [\psi^{\a i}] - \frac 23 [\l^{\a i }]~~~, ~~ \chi_{\a i}
\equiv  [\xi_{\a i}] ~~,~~ \l^{\a i j k}  \equiv [\psi^{\a i j k}] ~~,~~
\chi^{\a}{}_{ i j k}  \equiv [\xi^{\a}{}_{ i j k}] ~-~ i \frac 83 \pa^{\a  \b}
[{\bar \psi}_{\b i j k}] ~~~,
$$
$$
C^{i j k l} \equiv [C^{i j k l}] ~~,~~ L_{i j k l} \equiv [L_{i j k l}] ~~,~~
M^{i j} \equiv [K^{i j}] - \frac 32 [M^{i j}] ~~,~~ N \equiv [N] ~~,~~
$$
$$
K^{i j} \equiv [K^{i j}] ~+~ \frac 12 [M^{i j}] ~~,~~
G_{\a \b } \equiv [G_{\a \b}] ~+~  6 \pa_{\a  \b}[V] ~~,
$$
$$
V_{\a \b i j} \equiv [V_{\a \b i j}] ~-~ [A_{\a \b i j}] ~+~
4 \pa_{\a  \b}[L_{i j}] ~~,
$$
\begin{equation}
A_{\a \b i j} \equiv [A_{\a \b i j}] ~+~  4 \pa_{\a  \b}[L_{i j}] ~~,
\end{equation}
where we have adhered as closely as possible to the original notation. The
fields within the square brackets are the original definitions that appear
in reference \cite{HoSTT}.  Strictly speaking, these same definitions
ought to have been applied to the original N = 2, four dimensional
expressions.

The flat limit ``supersymmetry transformation laws'' (analogous to (1)
and (4)) can be found by combining the redefinitions above with results
in \cite{HoSTT}. For example, we find
\begin{equation}
D_{\a i} \varphi ~=~ \frac 12 \psi_{\a i} ~+~ \l_{\a i} ~~~,~~~
D_{\a i} \varphi_{j k} ~=~ C_{i (j| } \psi_{\a |k) } ~+~ 2 \l_{\a i j k } ~~~.
\end{equation}
It is simple exercise to write all such supersymmetry variations,
so we forego repeating them all here. The flat limit component
action when written in terms of the redefined fields above turns out
to be,
$$
 {\cal S}_{{\rm HP},{\rm N} = 4} ~=~  \int d^2 \s ~ [~
\varphi {}_\bo \varphi  ~+~ \varphi_{i j} {}_\bo \varphi^{i j} ~+~ i
\psi^{\a i} \pa_{\a \b} {\bar \psi}^{\b}{}_i {~~~~~~~~~~} {~~~~~~~~~~}
{~~~~~~~~~~~}
$$
$$ {~~~~~~~~~~~~~} +~ (~ \l^{\a i} \chi_{\a i} ~+~ {\rm h.}{\rm c.} ~)
{}~-~  2(~ \l^{\a i j k} \chi_{\a i j k} ~+~ {\rm h.}{\rm c.} ~)
$$
$$
{~~~~~~~~~~~~~~~} +~ \frac 1{18} N {\bar N} ~+~ \frac 38 (~K^{i j}
{\bar K}_{i j} -  M^{i j} {\bar M}_{i j} ~) - \frac 54 C^{ i j k l}
L_{ i j k l}
$$
\begin{equation}
{~~~~~~~~~~~~~~~~~} -~ \frac 1{36} G^{\a \b} G_{\a \b} ~-~ \frac 38 (~
 A^{ \a \b i j}  A_{ \a \b i j} -  V^{\a \b i j} V_{ \a \b i j} ~) ~]
{}~~~.
\end{equation}
In this expression, we have written some of the auxiliary fields utilizing
a matrix formulation.  For example we note, (with $G_{\a \b} = (G_{\b \a}
)^* $ as a consequence of the reduction from 4D)
\begin{equation}
G_{\a \b} ~\equiv~ C_{\a \b} {\cal B} ~+~ i (\g^3)_{\a \b} {\cal H} ~+~
 (\g^a)_{\a \b} {\cal V}_a ~~~,
\end{equation}
for real fields ${\cal B}$, ${\cal H}$ and ${\cal V}_a$.  Similar expressions
follow for $V_{\a \b i j}$ and $A_{\a \b i j}$.

\section{An N = 4 Superstring Question?}
\indent

Some time ago, Pernici and van Nieuwenhuizen \cite{vanN} (P-vN) gave
what was purported to be an action for the N = 4 superstring.  In fact
it was stated \cite{GHvN} that our off-shell construction must
describe this previous work. With the present investigation, however,
it is seen that this cannot be the case!  The models in reference
\cite{GHvN} involving twisted theories and be cannot identified with
the type of matter field representation used in reference \cite{vanN}.
Similarly, the HM-II N = 4 superstring discussed in the last section
also cannot be (at least not directly) identified with the P-vN
construction because the SU(2) under which the spin-1/2 fields form
a doublet is here used to put the four spin-0 fields into a 1s1p
representation. The corresponding fields are all singlets in the
P-vN work.  Even more surprisingly, careful study of the P-vN theory
shows that it is fundamentally different from any of the N = 4 models
discussed in this paper!

The first sign of this difference can be noted by comparing the
supergravity spectra of the theory in reference \cite{GaLO,GHvN} in
comparison with that given in \cite{vanN}. In the first of the
former, the 2D, N = 4 supergravity pre-potential is identified as
$V_{a i}{}^j$ a Lorentz-vector SU(2)-triplet superfield of
engineering dimensions minus three. This determines the field
content of conformal supergravity (as derived from superfield
theory) to be $e_a {}^m$, $\psi_a {}^{\a i}$ and $B_{a ~i}{}^j$.
The multiplet contain a graviton, complex SU(2) doublet gravitini
and the SU(2) gauge fields.  By way of comparison, P-vN have a
supergravity spectrum that consist of $e_a {}^m$, $\psi_a {}^{\a \hat
A i}$ and $B_{a ~i}{}^j$ where their gravitini satisfy a Majorana
condition.  This may be regarded as simply a redefinition of
the gravitini.  However, this also means that the corresponding
superspace description is based on a Majorana spinorial
covariant derivative $\nabla_{\a \hat A i}$. Appropriate
indices will thus appear in supertorsions.

In all of the 2D, N = 4 superstrings described in superspace,
the component level gravitini gauge covariant field strength
must be locally SU(2) covariant. The corresponding quantities
in the P-vN construction violate this condition!  Instead there
one of the connections associated with the non-linear $\s$-model
plays the role of $B_{a ~i}{}^j$. Since this particular covariantization
plays an absolutely critical role, it is not clear that the P-vN theory
can possess a limit where the $\s$-model manifold is flat and
the supergravity fields are non-trivial!  Closing this section,
we note that we have not in this work addressed the more general
question of the number of distinct local, N = 4 $\s$-models.  There
are many such rigid models (see for example \cite{DOGT,BLR}) that
can act as the starting point of such investigations.

\section{Conclusions and Summary}
\indent

One of the amusing points now obvious is that we have, in our previous
work \cite{GaLO,GHvN}, serendipitously discovered new N = 4 superstring
theories!  The original work on the N = 4 superstring \cite{11IJHS}
does not involve a twisted N = 4 supersymmetric matter multiplet. It is
the on-shell truncation of a hypermultiplet N = 4 superstring that was
considered in the first discussion of N = 4 superstrings.  The most
forceful way to see this is that the supersymmetry variation of the
scalar fields in twisted multiplets always includes the appearance
of the $\g^3$-matrix in at least one of the transformation laws. In the
corresponding variations of the 2D, N = 4 hypermultiplet, there is no
appearance of the $\g^3$-matrix.  Next we see that the twisted
N = 4 superstrings are models that realize (off-shell) the N = 4
superconformal algebra in a minimal fashion on a total of 16-16 fields.
Alternately, the HM-II, N = 4 superstring utilizes 40-40 component
to realize (off-shell) the N = 4 superconformal algebra.

In a future work, we will discuss issue of the construction and
characterization of superspace non-linear sigma models associated with
these three different N = 4
superstring actions. There remains the study in detail of the local form
of the component level actions and the explicit realization of superfield
local conformal transformations.  It also will be a topic of future study
for us to apply a BV type of quantization technique to these models and
study yet again the question of the critical dimension of N = 4 superstring
theory.  As we now have complete off-shell structures for all
of these theories, their quantization should follow as a straightforward
exercise.

As we have seen, there is an unexpected {\it {embarass de richesse}}
of off-shell N = 4 superstring theories. This is ironic since for over
a decade there did not exist even one off-shell formulation of the
N = 4 theory. In the following table, we summarize the features that
distinguish each of the N = 4 superstrings.
\begin{center}
\renewcommand\arraystretch{1.2}
\begin{tabular}{|c|c|c|c| }\hline
  $ $  & ${\rm TM-I}$  & ${\rm TM-II}$  & $~{\rm HM-II}~$
  \\ \hline \hline
${\rm Spin}-0 ~{\rm SU}(2) ~{\rm Rep.}$ & $4{\rm s}$  & $1{\rm s} 1{\rm p}$  &
$1{\rm s} 1{\rm p}$   \\ \hline
${\rm Off-shell} ~{\rm degrees} $ & ${\rm 16-16}$ & ${\rm 16-16}$ &
${\rm 40-40}$  \\ \hline
$(\g^3) ~{\rm in} ~{\rm Q-}{\rm trans.} ~{\rm law} $ & ${\rm yes}$ & ${\rm
yes}$ &
${\rm no}$  \\ \hline
\end{tabular}
\end{center}
\vskip.2in
\centerline{{\bf Table I}}
\noindent
These features will be apparent when the currents for the N = 4 models
are constructed using the techniques of superconformal field theory.
The presence of $(\g^3)$ implies that the left and right currents do
not appear in a totally isomorphic fashion.

Looking at the table above, also suggests that there exist one
more N = 4 superstring that has four SU(2)-singlet scalar fields and
{\underline {no}} $(\g^3)$ in its supersymmetry transformation laws
among the physical scalars and spinors.  The on-shell description of
this multiplet is given by
$$ D_{\a i} {\cal A} ~+~ C_{ i j} {\bar D}_{\a}{}^j {\bar {\cal B}} ~=~
{\bar D}_{\a}{}^i {\cal A} ~=~  {D}_{\a i} {\bar {\cal B}} ~=~ 0 ~~~,
$$
$$ D_{\a i} {\cal A} ~\equiv~ \varphi_{\a i} ~~~,~~~  {\bar D}_{\a}
{}^i {\bar {\cal B}} ~\equiv~ C^{i j}  \varphi_{\a j} ~~~,
$$
\begin{equation}
 D_{\a i}  \varphi_{\b j} ~=~ i 2  C_{i j} (\g^c)_{\a \b} \pa_c {\bar {\cal B}}
{}~~~,~~~ {\bar D}_{\a}{}^i  \varphi_{\b j} ~=~ i 2 \d_i {}^j  (\g^c)_{\a \b}
\pa_c {\cal A} ~~~.
\end{equation}
The invariant action without auxiliary fields (on-shell action) is given by,
\begin{equation}
{\cal S}_{ {\rm {HM-I,}~}{\rm {N = 4}~}} ~=~
\int d^2 \s ~[~ 2 {\bar {\cal A}} \bo {\cal A}
{}~+~ 2 {\bar {\cal B}} \bo {\cal B} ~+~ i {\bar {\varphi }}^{\a i}
(\g^c)_{\a \b} \pa_c  \varphi^{\b}{}_i ~~]~~~.
\end{equation}
Attempting an off-shell formulation raises the whole issue of the
``harmonic" hypermultiplet \cite{IV} which can be compactified to 2D.
This model with its {\it infinite} number of auxiliary fields gives us
the only known candidate for the off-shell description of the HM-I,
N = 4 scalar multiplet. It is not known how to describe the off-shell
representation with a finite number of auxiliary fields for this scalar
multiplet.  The real challenges are find out; (a.) if there is an
alternate off-shell formulation of the HM-I scalar multiplet with
a finite number of auxiliary fields and (b.) if the harmonic
formulation of the HM-I scalar multiplet can be coupled to minimal
2D, N = 4 supergravity to provide an off-shell description to the
HM-I, N = 4 superstring.

It is, perhaps, beneficial to use what we have learned and ask,
``Are there any more N = 4 superstrings?'' At this point, we do not
know how to find the answer to this one way or the other. There
are some possible variant representations of the hypermultiplet N
= 4 superstring. In the original work \cite{HoSTT}, it was
pointed out that the relaxed hypermultiplet could be formulated
with slightly different sets of auxiliary fields. Compactification to 2D
should preserve this possibility. So there is at least this ambiguity.
We have no principle for ruling out further N = 4 superstrings.

The final interesting point raised by our title may have a resolution
in terms of extending the notion of mirror symmetry to act between a
hyper Kahler manifold and a quaternionic manifold. This is an exciting
prospect. But for now we are simply left with the puzzle of our title.

$$ {~~~} $$

\noindent {\bf{Acknowledgement} }
\indent \newline
S. J. G. wishes to acknowledge the hospitality of Drs. S. V. Ketov,
O. Lechtenfeld and the Institut f\" ur Theoretische Physik of
Universit\" at Hannover extended during the period of this work.
Useful comments from Martin Ro\v cek are also acknowledged.

$$ {~~~~~~~}$$

\newpage


\begin{thebibliography}{66}

\bibitem{GaLO} S. James Gates, Jr., Liang Lu and R. Oerter, Phys. Lett. 218B
(1989) 33.

\bibitem{GaHR} S. James Gates, Jr., C. M. Hull and M. Ro\v cek, Nucl. Phys.
B248 (1984) 157.

\bibitem{KKL} E. Kiritis, C. Kounnas and D. L\" ust, {\it
{Superstring Gravitational Wave Backgrounds with Spacetime Supersymmetry}},
preprint, CERN-TH.7218/94, HUB-IEP-94/5, LPTENS-94/10, hep-th/9404114.

\bibitem{11IJHS} M. Admello, L. Brink, A. D'Adda, R. D'Auria, E. Napolitano,
S. Sciuto, E. Del Giudice, P. Di Vecchia, S. Ferrara, F. Gliozzi, R.
Musto, R. Pattorino, and J. Schwarz, Nucl. Phys. 114 (1976) 297.

\bibitem{vanN} P. van Nieuwenhuizen, Int. J. Mod. Phys. A1, (1986), 155; M.
Pernici and P. van Nieuwenhuizen, Phys. Lett. 169B (1986) 381.

\bibitem{sieg} W. Siegel, Phys. Rev. Lett. 69 (1992) 1493;  Phys. Rev.
D46 (1992) 3235; ibid. D47 (1993) 2504, ibid. 2512.

\bibitem{GHvN} S. James Gates, Jr., Y. Hassoun and P. van Nieuwenhuizen,
Nucl. Phys. B317 (1989) 302.

\bibitem{BrkGT} R. Brooks and S. James Gates, Jr., {\it {Extended Supersymmetry
and Super-BF Gauge Theory}}, M.I.T. preprint CTP-2339, UMD preprint
UMDEPP 95-07.

\bibitem{SG1} S. James Gates, Jr. and W. Siegel, Nucl. Phys. B187 (1981) 389.

\bibitem{IVK} E.A. Ivanov and S.O. Krivonos, J. Phys. A17 (1984) L671;
idem. Theor. Math. Phys. 63 (1985) 477.

\bibitem{FrdAlGat} D.Z. Freedman and L. Alvarez-Gaum\' e, Commun. Math.
Phys. 91 (1983) 87;  S. James Gates, Jr., Nucl. Phys. B238 (1984) 349;
A. Jourjine,  Nucl. Phys. B236 (1984) 181.

\bibitem{DOGT} B.B. Deo and S. James Gates, Jr., Nucl. Phys. B254 (1985) 187.

\bibitem{HoSTT} P.S. Howe, K.S.Stelle, and P.K. Townsend, Nucl. Phys.
B214 (1983) 519; S.V.Ketov, J. Mod Phys. A3, (1988) 703; idem., Fort.
der Physik, 36 (1988) 361.

\bibitem{WePK} P.S. Howe, K.S.Stelle, and P. West, Phys. Lett. 124B
(1983) 55; P. West, in {\it Proceedings of the Shelter Island II Conference
on Quantum Field Theory and Fundamental Problems of Physics}, eds. R.Jackiw,
N.Khuri, S. Weinberg and E. Witten, M.I.T. Press (1993).

\bibitem{IV} A. Galperin, E. Ivanov, V. Ogievetsky and E. Sokatchev,
JETP Lett. 40 (1984) 912; A. Galperin, E. Ivanov, S. Kalitzin,
V. Ogievetsky and E. Sokatchev, Class. Quant. Grav. 1 (1984) 469.

\bibitem{BLR} T. Buscher, U. Lindstrom and M. Ro\v cek, Phys. Lett. 202B
(1988),  U. Lindstrom and M. Ro\v cek, Commun. Math. Phys. 115 (1988) 21.

\end{thebibliography}
\end{document}